\documentstyle[12pt,graphicx,amsmath,tabularx,multirow]{article}
\newlength{\pubnumber} 

\catcode`\@=11
\@addtoreset{equation}{section}

\def\section{\@startsection{section}{1}{\z@}{3.5ex plus 1ex minus .2ex}
 {2.3ex plus .2ex}{\large\bf}}
\def\subsection{\@startsection{subsection}{2}{\z@}{2.3ex plus .2ex}
 {2.3ex plus .2ex}{\bf}}

\newcommand{\ba}{\begin{eqnarray}}
\newcommand{\ea}{\end{eqnarray}}
\begin{document}

\begin{titlepage}
\samepage{
\setcounter{page}{1}
\rightline{LTH--953}
\rightline{July 2012}

\vfill
\begin{center}
 {\Large \bf
String Derived Exophobic $SU(6)\times SU(2)$ GUTs
}
\vspace{1cm}
\vfill {\large
Laura Bernard$^{1}$,
Alon E Faraggi$^{2}$\\
\vspace{.1in}
Ivan Glasser$^{1}$,
John Rizos$^{3}$ and
Hasan Sonmez$^{2}$
}\\
\vspace{1cm}
{\it $^{1}$ Centre de Physique Th\'eorique,
             Ecole Polytechnique,
         F--91128 Palaiseau, France\\}
\vspace{.05in}
{\it $^{2}$ Dept.\ of Mathematical Sciences,
             University of Liverpool,
         Liverpool L69 7ZL, UK\\}
\vspace{.05in}
{\it $^{3}$ Department of Physics,
              University of Ioannina, GR45110 Ioannina, Greece\\}
\vspace{.025in}
\end{center}
\vfill
\begin{abstract}
With the apparent discovery of the Higgs boson, the Standard Model
has been confirmed as the theory accounting for all sub--atomic
phenomena. This observation lends further credence to the
perturbative unification in Grand Unified Theories (GUTs) and string theories.
The free fermionic formalism yielded fertile ground for
the construction of quasi--realistic heterotic--string models,
which correspond to toroidal $Z_2\times Z_2$ orbifold compactifications.
In this paper we study a new class of heterotic--string models
in which the GUT group is $SU(6)\times SU(2)$ at the string
level. We use our recently developed fishing algorithm to
extract an example of a three generation $SU(6)\times SU(2)$
GUT model. We explore the phenomenology of the model and
show that it contains the required symmetry breaking Higgs
representations. We show that the model admits flat directions
that produce a Yukawa coupling for a single family. The
novel feature of the $SU(6)\times SU(2)$ string GUT models
is that they produce an additional family universal anomaly
free $U(1)$ symmetry, and 
may remain unbroken below the string scale.
The massless spectrum of the model is free
of exotic states.

\noindent

\end{abstract}
\smallskip}
\end{titlepage}

\setcounter{footnote}{0}

\def\beq{\begin{equation}}
\def\eeq{\end{equation}}
\def\beqn{\begin{eqnarray}}
\def\eeqn{\end{eqnarray}}

\def\no{\noindent }
\def\nolabel{\nonumber }
\def\ie{{\it i.e.}}
\def\eg{{\it e.g.}}
\def\half{{\textstyle{1\over 2}}}
\def\third{{\textstyle {1\over3}}}
\def\quarter{{\textstyle {1\over4}}}
\def\sixth{{\textstyle {1\over6}}}
\def\m{{\tt -}}
\def\p{{\tt +}}

\def\Tr{{\rm Tr}\, }
\def\tr{{\rm tr}\, }

\def\slash#1{#1\hskip-6pt/\hskip6pt}
\def\slk{\slash{k}}
\def\GeV{\,{\rm GeV}}
\def\TeV{\,{\rm TeV}}
\def\y{\,{\rm y}}
\def\SM{Standard--Model }
\def\SUSY{supersymmetry }
\def\SSSM{supersymmetric standard model}
\def\vev#1{\left\langle #1\right\rangle}
\def\l{\langle}
\def\r{\rangle}
\def\o#1{\frac{1}{#1}}

\def\Htw{{\tilde H}}
\def\chibar{{\overline{\chi}}}
\def\qbar{{\overline{q}}}
\def\ibar{{\overline{\imath}}}
\def\jbar{{\overline{\jmath}}}
\def\Hbar{{\overline{H}}}
\def\Qbar{{\overline{Q}}}
\def\abar{{\overline{a}}}
\def\alphabar{{\overline{\alpha}}}
\def\betabar{{\overline{\beta}}}
\def\tautwo{{ \tau_2 }}
\def\thetatwo{{ \vartheta_2 }}
\def\thetathree{{ \vartheta_3 }}
\def\thetafour{{ \vartheta_4 }}
\def\ttwo{{\vartheta_2}}
\def\tthree{{\vartheta_3}}
\def\tfour{{\vartheta_4}}
\def\ti{{\vartheta_i}}
\def\tj{{\vartheta_j}}
\def\tk{{\vartheta_k}}
\def\calF{{\cal F}}
\def\smallmatrix#1#2#3#4{{ {{#1}~{#2}\choose{#3}~{#4}} }}
\def\ab{{\alpha\beta}}
\def\Minv{{ (M^{-1}_\ab)_{ij} }}
\def\bone{{\bf 1}}
\def\ii{{(i)}}
\def\V{{\bf V}}
\def\N{{\bf N}}

\def\b{{\bf b}}
\def\S{{\bf S}}
\def\X{{\bf X}}
\def\I{{\bf I}}
\def\mb{{\mathbf b}}
\def\mS{{\mathbf S}}
\def\mX{{\mathbf X}}
\def\mI{{\mathbf I}}
\def\balpha{{\mathbf \alpha}}
\def\bbeta{{\mathbf \beta}}
\def\bgamma{{\mathbf \gamma}}
\def\bxi{{\mathbf \xi}}

\def\t#1#2{{ \Theta\left\lbrack \matrix{ {#1}\cr {#2}\cr }\right\rbrack }}
\def\C#1#2{{ C\left\lbrack \matrix{ {#1}\cr {#2}\cr }\right\rbrack }}
\def\tp#1#2{{ \Theta'\left\lbrack \matrix{ {#1}\cr {#2}\cr }\right\rbrack }}
\def\tpp#1#2{{ \Theta''\left\lbrack \matrix{ {#1}\cr {#2}\cr }\right\rbrack }}
\def\l{\langle}
\def\r{\rangle}
\newcommand{\cc}[2]{c{#1\atopwithdelims[]#2}}
\newcommand{\nn}{\nonumber}


\def\inbar{\,\vrule height1.5ex width.4pt depth0pt}

\def\IC{\relax\hbox{$\inbar\kern-.3em{\rm C}$}}
\def\IQ{\relax\hbox{$\inbar\kern-.3em{\rm Q}$}}
\def\IR{\relax{\rm I\kern-.18em R}}
 \font\cmss=cmss10 \font\cmsss=cmss10 at 7pt
\def\IZ{\relax\ifmmode\mathchoice
 {\hbox{\cmss Z\kern-.4em Z}}{\hbox{\cmss Z\kern-.4em Z}}
 {\lower.9pt\hbox{\cmsss Z\kern-.4em Z}}
 {\lower1.2pt\hbox{\cmsss Z\kern-.4em Z}}\else{\cmss Z\kern-.4em Z}\fi}

\def\AEF{A.E. Faraggi}
\def\JHEP#1#2#3{{\it JHEP}\/ {\bf #1} (#2) #3}
\def\NPB#1#2#3{{\it Nucl.\ Phys.}\/ {\bf B#1} (#2) #3}
\def\PLB#1#2#3{{\it Phys.\ Lett.}\/ {\bf B#1} (#2) #3}
\def\PRD#1#2#3{{\it Phys.\ Rev.}\/ {\bf D#1} (#2) #3}
\def\PRL#1#2#3{{\it Phys.\ Rev.\ Lett.}\/ {\bf #1} (#2) #3}
\def\PRT#1#2#3{{\it Phys.\ Rep.}\/ {\bf#1} (#2) #3}
\def\MODA#1#2#3{{\it Mod.\ Phys.\ Lett.}\/ {\bf A#1} (#2) #3}
\def\IJMP#1#2#3{{\it Int.\ J.\ Mod.\ Phys.}\/ {\bf A#1} (#2) #3}
\def\nuvc#1#2#3{{\it Nuovo Cimento}\/ {\bf #1A} (#2) #3}
\def\RPP#1#2#3{{\it Rept.\ Prog.\ Phys.}\/ {\bf #1} (#2) #3}
\def\EJP#1#2#3{{\it Eur.\ Phys.\ Jour.}\/ {\bf C#1} (#2) #3}
\def\etal{{\it et al\/}}

\hyphenation{su-per-sym-met-ric non-su-per-sym-met-ric}
\hyphenation{space-time-super-sym-met-ric}
\hyphenation{mod-u-lar mod-u-lar--in-var-i-ant}


\setcounter{footnote}{0}
\section{Introduction}

With the imminent discovery of the Higgs boson looming another
piece in the particle puzzle falls into place. Confirmation
of its scalar identity is another vital goal, which may,
however, require a precision machine to study its properties
in greater detail. The discovery of additional scalar particles
has similarly eluded experiments to date and pushes
the scale of supersymmetric particles, possibly beyond the
LHC scale. 
Nevertheless, the discovery of a fundamental Higgs
boson lends further credence to the prevailing picture
of unifying the observed gauge and matter structure
in a Grand Unified Theory, or, ultimately in string theory.

This unification in turn gives rise to new puzzles.
Among the most perplexing is that of proton longevity.
Indeed, in the Standard Model proton stability is only
guaranteed due to the existence of accidental global
symmetries at the renormalisable level. Proton
instability becomes especially acute in theories
of quantum gravity that are not expected to
preserve global symmetries. In the context of
string constructions it was
proposed that the proton stability problem may
indicate the existence of an additional $U(1)$ symmetry
at a comparatively low scale that serves as a proton
lifeguard \cite{plg}. The caveat, however, is that
other phenomenological requirements constrain
possible $U(1)$ symmetries to be broken
at the high, or intermediate, scale. Thus,
failing to provide adequate suppression of
proton decay mediating operators. In general,
it is found that keeping an additional $U(1)$ symmetry
unbroken in string constructions is highly nontrivial.

To explore such questions in detail one must construct
phenomenological string models. The free fermionic formulation
\cite{fff} of the heterotic string provided a fertile framework
to study quasi--realistic string vacua. Three generation
models in this construction preserve the $SO(10)$ embedding
of the Standard Model states \cite{fsu5,slm,alr,lrs,su421}.
The existence of models in this class which produce solely the
spectrum of the Minimal Supersymmetric Standard Model
has been further demonstrated. The free fermionic models
correspond to $Z_2\times Z_2$ orbifold compactifications
with discrete Wilson lines \cite{z2z21,z2z22}. The $SO(10)$ symmetry is
broken directly at the string level to one of its subgroups,
which included to date:
the flipped $SU(5)$ (FSU5) \cite{fsu5};
the standard--like models (SLM) \cite{slm};
the Pati--Salam models (PS) \cite{alr,acfkr};
the left--right symmetric models \cite{lrs};
and the $SU(4)\times SU(2) \times U(1)$ models (SU421)
\cite{su421}. Of those, it was shown that the
SU421 models in this class do not produce realistic
spectra \cite{su421}.

The proposition of a proton lifeguard extra $U(1)$ symmetry
in these models was studied in ref. \cite{plg}.
The $U(1)$ symmetry that can play this role
in the free fermionic models is a combination of $U(1)_{B-L}$,
which is embedded in $SO(10)$ and a family universal
combination of three $U(1)$s in the Cartan subalgebra
of the observable $E_8$, and which are external to
$SO(10)$. We shall refer to this $U(1)$ combination
as $U(1)_\zeta=U(1)_1+U(1)_2+U(1)_3$.
The caveat, however, is that in the FSU5, SLM and PS free fermionic
modes the $U(1)_\zeta$ combination is anomalous and hence must be
broken near the string scale. In the LRS string models it is anomaly
free, but has mixed $SU(2)_L^2\times U(1)_\zeta$ anomalies \cite{viraf1}.
This necessitates the introduction of additional lepton doublets
to render the spectrum anomaly free, which requires also
additional colour triplets to facilitate gauge coupling unification
\cite{viraf2}.

In this paper we study a new class of string models that can produce
the desired proton lifeguard $U(1)$ combination. Particularly,
in these models $U(1)_\zeta$ is anomaly free. The models
under consideration possess an $SU(6)\times SU(2)$ unbroken
symmetry at the string level and arise from the breaking of $E_6$
to this maximal subgroup. Thus, the $U(1)_\zeta$ combination is
anomaly free by virtue of its $SU(6)$ embedding. It should be
noted that the viability of an extra $U(1)$ down to low scales
may result in additional phenomenological issues that may need
to be addressed. For example, the augmentation of the spectrum
by lepton doublets in the model of ref. \cite{viraf1} requires
additional colour triplets that may, a priori, lead to proton
decay mediating operators. To construct a full model with a
viable $U(1)$ down to the low scale is beyond the scope of
this paper. Therefore, the possibility also exist to break
the symmetry directly to the Standard Model at the high scale.
We present an example of a three generation free fermionic
$SU(6)\times SU(2)$ GUT model and explore its phenomenological viability.

Our paper is organised as follows: in section \ref{fieldsu62}
we review the field theory structure of the $SU(6)\times SU(2)$ models. In
section \ref{analysis} we discuss the construction of $SU(6)\times SU(2)$
heterotic--string models and present an exemplary three generation
free fermionic model. In section \ref{pheno} we present the cubic
level superpotential of the model and explore its phenomenology.
Section \ref{conclusion} concludes the paper.

\section{The supersymmetric $SU(6)\times SU(2)$
Models}\label{fieldsu62}

In this section we briefly summarise the field theory content of the
$SU(6)\times SU(2)$ models.
This gauge group is a maximal subgroup of $E_6$.
The matter states of the model are obtained therefore from the $27$
representation
of $E_6$ decomposed under the $SU(6)\times SU(2)$ subgroup. We note that we have
two options of embedding the Standard Model into this subgroup. Namely, we
can choose the electroweak $SU(2)_L$ to reside inside $SU(6)$ or, alternatively,
we can identify it with the $SU(2)$ which is orthogonal to $SU(6)$.
The field theory model building and symmetry breaking patterns
of the $SU(6)\times SU(2)$ GUT models were discussed in ref.
\cite{dh,rt}, as well as the possibility of preserving an unbroken
$U(1)$ symmetry to low scales \cite{kim}.
The advantage of having an $SU(6)$ GUT in respect to proton
lifetime \cite{raby}, as well as the flexibility it affords in regard
to gauge coupling unification \cite{rt}, has also been examined.
The novelty in our
paper is the derivation of a string model that can realise the
$SU(6)\times SU(2)$ GUT model. Further details of the
field theory construction can be found in \cite{rt,kim},
and will be further explored in a future publication.
The non--trivial aspect
From the point of view of the string
model building is that $U(1)_\zeta$ is anomaly free. Turning to the
field theory content, the $27$ representation of $E_6$ decomposes as
\beq
27=(15,1)+ ({\bar 6},2)
\label{27undersu62}
\eeq
Our string models will be constructed in the following as an enhancement of the
Pati--Salam heterotic string models. It is instructive to note the decomposition
in the Pati--Salam and Standard--Model subgroups.
In this regard the $27$ representation
of $E_6$ decomposes under $SO(10)\times U(1)$ as
\beq
27=(16,-{1\over2})+(10,1)+(1,-2),
\label{27underso101}
\eeq
where the $16$ and $10$ are the spinorial and vectorial
representations of $SO(10)$, respectively. Under the Pati--Salam subgroup
these decompose into representations of
$SU(4)\times SU(2)_R\times SU(2)_L\times U(1)_\zeta$
as
\beqn
(16,-{1\over2}) & = & F_L+ F_R =~ (4,2,1,-{1\over2}) +
                                ({\bar 4}, 1,2,-{1\over2})\nonumber\\
(10,~~1) & = & {\cal D} ~+ ~\hbox{h} ~=~ (6,1,1,~~1) ~+ (1,2,2,~~1) \nonumber\\
(1,-2)   & = & ~~~~{\cal S}~~~~~ =~ (1,1,1,-2).\nonumber
\eeqn
The decomposition of the $27$ under
$SU(3)_C\times SU(2)_L\times U(1)_{C}\times U(1)_L\times U(1)_\zeta$
is shown in table \ref{27dusu3211}.

\begin{table}[!h]
\noindent
{\small
\openup\jot
\begin{center}
\begin{tabular}{|l|cc|c|c|c|}
\hline
Field&$\hphantom{\times}
  SU(3)_C$&$\times SU(2)_L $&$U(1)_{C}$&${U(1)}_L$&${U(1)}_{\zeta}$\\[1ex]
\hline
$Q$  &$3$      &$2$&$+\frac{1}{2} $&$\hphantom{+}0$&$-\frac{1}{2}$\\[1ex]
$L$  &$1$      &$2$&$-\frac{3}{2} $&$\hphantom{+}0$&$-\frac{1}{2}$\\[1ex]
$u$  &$\bar{3}$&$1$&$-\frac{1}{2}$ &$-1$&$-\frac{1}{2}$\\[1ex]
$d$  &$\bar{3}$&$1$&$-\frac{1}{2}$ &$+1$&$-\frac{1}{2}$\\[1ex]
$e$  &$1$      &$1$&$+\frac{3}{2}$ &$+1$&$-\frac{1}{2}$\\[1ex]
$N$&$1$        &$1$&$+\frac{3}{2}$ &$-1$&$-\frac{1}{2}$\\[1ex]
\hline
$h^u$    &$1$      &$2$&$\hphantom{+}0$&$+1$&$+1$\\[1ex]
$h^d$    &$1$      &$2$&$\hphantom{+}0$&$-1$&$+1$\\[1ex]
$D$      &$3$      &$1$&$-1$&$\hphantom{+}0$&$+1$\\[1ex]
$\bar{D}$&$\bar{3}$&$1$&$+1$&$\hphantom{+}0$&$+1$\\[1ex]
\hline
${\cal S}$&$1$&$1$&$\hphantom{+}0$&$\hphantom{+}0$&$-2$\\[1ex]
\hline
\end{tabular}
\end{center}
}
\caption{\label{27dusu3211}\it
Decomposition of the $27$ representation of $E_6$ under
$SU(3)_C\times SU(2)_L\times U(1)_C\times U(1)_L\times U(1)_\zeta$, where
$U(1)_C={3\over 2} U(1)_{B-L}$ is the Baryon minus Lepton number and
$U(1)_L= 2 U(1)_{3R}$ is the diagonal generator of $SU(2)_R$.  }
\end{table}

If we choose the electroweak $SU(2)_L$ gauge group to be the one
external to $SU(6)$, then
the Standard Model matter and Higgs representations are embedded in
\beqn
{\cal F}_L^i & = & ({\bar 6}, 2)^i = F_L^i + \hbox{h}^i =
(Q + L +h^u+h^d)^i \label{fli62}\\
{\cal F}_R^i & = & (15,1)^i = F_R^i + {\cal D}^i + {\cal S}^i =
(u+ d + e + N+ D + {\bar D}
+ {\cal S})^i\label{fri151}
\eeqn
where the index $i=1,2,3$ is a generation index. The weak
hypercharge combination is given
by\footnote{$U(1)_C=3/2U(1)_{B-L}~;~U(1)_L=U(1)_{T_{3_R}}$.}

\beq
U(1)_Y= {1\over 3} U(1)_C + {1\over 2} U(1)_L
\label{weakhypercharge}
\eeq
In this case the heavy Higgs states in the model are in the $(15,1)$ and
$(\overline{15},1)$ representations of $SU(6)\times SU(2)$.
We need two pairs, $H+\overline{H}$
and ${\cal H}+\overline{\cal H}$
to generate the symmetry breaking down to the Standard Model
gauge group $SU(3)_C\times SU(2)_L\times U(1)_Y$. The
$H+\overline{H}$ fields obtain VEVs in the direction of the
$N$ and $\overline{N}$ components. These VEVs breaks
the gauge symmetry to $SU(3)\times SU(2)\times U(1)_Y\times U(1)_{Z^\prime}$.
The ${\cal H}+\overline{\cal H}$ fields get VEVs in the direction
of the ${\cal S}$ and $\overline{\cal S}$ components, which break the
gauge symmetry down to the Standard Model. The first set of VEVs
has to be sufficiently large to generate a seesaw mechanism and
light neutrino masses, whereas the second set of VEVs can be 
at a lower scale. 
The fields in the $({\bar 6},2)$ representation of
$SU(6)\times SU(2)$ contain the electroweak Higgs bi--doublets that are
required to break the $SU(2)_L\times U(1)_Y$ down to $U(1)_{\rm e.m.}$.

The most general cubic level superpotential involving the $SU(6)\times SU(2)$
charged fields is given by
\beqn
W=& & \lambda^1_{ijk}{\cal F}_L^i{\cal F}_R^j{\cal F}_L^k+
  \lambda^2_{ij}{\cal F}_R^i{\cal F}_R^j{H}+
  \lambda^3_{ij}{\cal F}_R^i{\cal F}_R^j{\cal H}+
  \nonumber\\  & &
  \lambda^4_{ij}{\cal F}_L^i{\cal F}_L^j{H}+
  \lambda^5_{ij}{\cal F}_L^i{\cal F}_L^j{\cal H}+
  \lambda^6 HH{\cal H}+
  {\lambda}^7 \overline{H}\overline{H}\overline{\cal H}
\label{superpot1}
\eeqn
We note that some of these couplings may need to vanish for
viable phenomenological realisation.
The first term generates the fermion Yukawa couplings to the
electroweak Higgs fields.
The second and third terms
may generate mass terms for some of the colour triplet fields.
The sixth and seventh term generate the heavy Higgs superpotential.
The fourth term couples the
VEV of ${\cal S}$ to the electroweak bi--doublets. Hence, if this
VEV is at the low scale, it provides an explanation for the
suppression of the supersymmetric $\mu$--term. The scale
of the $\mu$--term is associated then with the scale of $U(1)_{Z^\prime}$
breaking. The neutrino seesaw mechanism
may be generated by coupling the right--handed neutrino components to light
$SU(6)\times SU(2)$ singlet fields, {\it i.e.}
\beq
\lambda^5_{ij} {\cal F}_R^i \overline{H} \phi^j+
\lambda^6_{ijk}\phi^i\phi^j\phi^k
\eeq
This scenario therefore requires the introduction of three additional
$SU(6)\times SU(2)$ singlets. Alternatively, the seesaw mass terms may be
obtained from dimension five terms
\beq
\lambda^5_{ij} {\cal F}_R^i{\cal F}_R^j\overline{H}\overline{H}
\label{seesawtermd4}
\eeq
Turning to the proton decay mediating operators, we note that the terms
arising from the quartic $16^4$ as well as those arise from dimension
four operators in the MSSM are forbidden in this model by $U(1)_\zeta$.
This is because we identified the Standard Model states as arising solely
from the $16$ representations of $SO(10)$ and all these states carry the
$U(1)_\zeta$ charge, as dictated by the $E_6$ charge assignment.
Therefore, both the dimension four and five operators are
forbidden. Preservation of an unbroken $U(1)_{Z^\prime}$, which is a
combination of $U(1)_{B-L}$ and $U(1)_\zeta$, then guarantees
that the proton decay mediating operators, as well as the Higgs
$\mu$--term are adequately suppressed
in this model. A more elaborate field theory analysis of the
model will be presented elsewhere.

The alternative to embedding the $SU(2)_R$ in $SU(6)$, as outlined above,
is to embed $SU(2)_L$ in $SU(6)$ and $SU(2)_R$ external to it.
In this case the Standard Model matter and Higgs states are embedded in
\beqn
{\cal F}_R^i & = & ({\bar 6}, 2)^i = F_R^i + \hbox{h}^i =
(u+ d + e + N+h^u+h^d)^i \label{fri62}\\
{\cal F}_L^i & = & (15,1)^i = F_L^i + {\cal D} + {\cal S}^i =
(Q + L + D + {\bar D}
+ {\cal S})^i\label{fli151}
\eeqn
where the index $i=1,2,3$ is a generation index. It is seen that the
effect is to flip the $F_L^i$ and $F_R^i$ representations
between the two representations in (\ref{fli62}) and (\ref{fri151}).
The weak hypercharge combination is still given by eq. (\ref{weakhypercharge})
in terms of the generators of the Cartan Subalgebra of
the observable $E_8$. The consequence is that with the assignment in
(\ref{fri62}) and (\ref{fli151}) two states of
the heavy Higgs representations are in the
$({\bar 6},2)$ and $(6,2)$, denoted by $H$ and $\overline{H}$,
which contain the VEVs in the direction of the $N$ and $\overline{N}$
components. The two other heavy Higgs states are, as before, in the
$(15,1)$ and $(\overline{15},1)$, denoted by ${\cal H}+\overline{\cal H}$,
contain the VEVs in the direction of the
${\cal S}$ and $\overline{\cal S}$ components.
Hence, in this case
the general cubic level superpotential involving the $SU(6)\times SU(2)_R$
charged fields is given by
\beq
W= \lambda^1_{ijk}{\cal F}_R^i{\cal F}_L^j{\cal F}_R^k+
  \lambda^2_{ij}{\cal F}_R^i{\cal F}_R^j{\cal H}+
  \lambda^3_{ij}{\cal F}_L^i{\cal F}_L^j{\cal H}+
  \lambda^4 HH{\cal H}+
  {\lambda}^5 \overline{H}\overline{H}\overline{\cal H}
\label{superpot2}
\eeq
The terms in
(\ref{superpot2}) are given in terms of the assignment in
(\ref{fri62}) and (\ref{fli151}). The seesaw
term can be taken from (\ref{seesawtermd4}), but with the
modified field assignment in (\ref{fri62}) and (\ref{fli151}).

We note that in the $SU(6)\times SU(2)_L$ string models the
heavy Higgs states may arise solely from the untwisted
sector, whereas the $SU(6)\times SU(2)_R$ string models
also require heavy string states that arise from twisted sectors.

\section{$SU(6)\times SU(2)$ Heterotic--String GUT
Models}\label{analysis}

In this section we present a three generation
$SU(6)\times SU(2)$ string GUT model.
The model is constructed in the free fermionic formulation by using
the classification method developed in refs. \cite{gkr,fknr,fkr, acfkr}.
The set of basis vectors is identical to the one used in the classification
of the Pati--Salam heterotic--string models \cite{acfkr}. The difference
between the two cases is that in \cite{acfkr} the vector bosons that enhance
the observable $SO(16)$ gauge symmetry to $E_8$ are projected out,
whereas here the are retained. Therefore, the $SU(6)\times SU(2)$
gauge symmetry is obtained here as enhancement of the
$SU(4)\times SU(2)_L\times SU(2)_R\times U(1)_1\times U(1)_2\times U(1)_3 $
observable gauge symmetry arising from the untwisted sector,
to $SU(6)\times SU(2)_L\times U(1)_1^\prime \times U(1)_2^\prime$.
We then follow a fishing algorithm in which we generate random
choices of GSO projection coefficients. We fish out the choices that
satisfy the enhancement criterion and additional phenomenological
criteria, like the existence of three generations and absence of exotics.

In the free fermionic formulation the 4--dimensional heterotic--string, in the
light-cone gauge, is described
by $20$ left--moving  and $44$ right--moving real fermions.
The models are constructed by choosing
different phases picked up by   fermions ($f_A, A=1,\dots,44$) when transported
around non--contractible loops of the world--sheet torus.
Each model corresponds to a choice of fermion phases consistent with
modular invariance
that can be generated by a set of  basis vectors $v_i,i=1,\dots,n$
$$v_i=\left\{\alpha_i({{\bar{f}_1}}),\alpha_i(f_{2}),\alpha_i(f_{3}))\dots\right\}$$
describing the transformation  properties of each fermion
\begin{equation}
f_A\to -e^{i\pi\alpha_i(f_A)}\ f_A, \ , A=1,\dots,44
\end{equation}
The basis vectors span a space $\Xi$ which consists of $2^N$ sectors that give
rise to the string spectrum. Each sector is given by
\begin{equation}
\xi = \sum N_i v_i,\ \  N_i =0,1
\end{equation}
The spectrum is truncated by a generalised GSO projection whose action on a
string state  $|S>$ is
\begin{equation}\label{eq:gso}
e^{i\pi v_i\cdot F_S} |S> = \delta_{S}\ \cc{S}{v_i} |S>,
\end{equation}
where $F_S$ is the fermion number operator and $\delta_{S}=\pm1$ is the
space--time spin statistics index.
Different sets of projection coefficients $\cc{S}{v_i}=\pm1$ consistent with
modular invariance give
rise to different models. Summarising: a model can be defined uniquely by a set
of basis vectors $v_i,i=1,\dots,n$
and a set of $2^{N(N-1)/2}$ independent projections coefficients
$\cc{v_i}{v_j}, i>j$.

The free fermions in the light-cone gauge in the usual notation are:
$\psi^\mu, \chi^i,y^i, \omega^i, i=1,\dots,6$ (left--movers) and
$\bar{y}^i,\bar{\omega}^i, i=1,\dots,6$,
$\psi^A, A=1,\dots,5$, $\bar{\eta}^B, B=1,2,3$, $\bar{\phi}^\alpha,
\alpha=1,\ldots,8$ (right--movers).
The class of models we investigate, is generated by a set of
thirteen basis vectors
$
B=\{v_1,v_2,\dots,v_{13}\},
$
where
\begin{eqnarray}
v_1=1&=&\{\psi^\mu,\
\chi^{1,\dots,6},y^{1,\dots,6}, \omega^{1,\dots,6}| \nonumber\\
& & ~~~\bar{y}^{1,\dots,6},\bar{\omega}^{1,\dots,6},
\bar{\eta}^{1,2,3},
\bar{\psi}^{1,\dots,5},\bar{\phi}^{1,\dots,8}\},\nonumber\\
v_2=S&=&\{\psi^\mu,\chi^{1,\dots,6}\},\nonumber\\
v_{2+i}=e_i&=&\{y^{i},\omega^{i}|\bar{y}^i,\bar{\omega}^i\}, \
i=1,\dots,6,\nonumber\\
v_{9}=b_1&=&\{\chi^{34},\chi^{56},y^{34},y^{56}|\bar{y}^{34},
\bar{y}^{56},\bar{\eta}^1,\bar{\psi}^{1,\dots,5}\},\label{basis}\\
v_{10}=b_2&=&\{\chi^{12},\chi^{56},y^{12},y^{56}|\bar{y}^{12},
\bar{y}^{56},\bar{\eta}^2,\bar{\psi}^{1,\dots,5}\},\nonumber\\
v_{11}=z_1&=&\{\bar{\phi}^{1,\dots,4}\},\nonumber\\
v_{12}=z_2&=&\{\bar{\phi}^{5,\dots,8}\},\nonumber\\
v_{13}=\alpha &=& \{\bar{\psi}^{4,5},\bar{\phi}^{1,2}\}.\nonumber
\end{eqnarray}
The first twelve vectors in this set are identical to those used in
\cite{fknr,fkr}
for the classification of $Z_2\times Z_2$ heterotic--string models
with an $SO(10)$
GUT group. The basis vector
$v_{13}$ is the vector that breaks the $SO(10)$ GUT symmetry to
$SO(6)\times SO(4)$. The second ingredient that is needed to
define the string vacuum
are the Generalised GSO (GGSO)
projection coefficients that appear in the one--loop partition function,
$\cc{v_i}{v_j}$, spanning a $13\times 13$ matrix.
Only the elements with $i>j$ are
independent, and the others are fixed by modular invariance.
A priori there are therefore 78 independent coefficients corresponding
to $2^{78}$ distinct string vacua. Eleven coefficients
are fixed by requiring that the models possess $N=1$ supersymmetry.
An explicit choice of GGSO projection coefficients that produces
a model with $SU(6)\times SU(2)$ GUT group is given
by the following GGSO coefficients matrix :

\beq \label{BigMatrix}  (v_i|v_j)\ \ =\ \ \bordermatrix{
& 1& S&e_1&e_2&e_3&e_4&e_5&e_6&b_1&b_2&z_1&z_2&\alpha\cr
1  & 1& 1& 1& 1& 1& 1& 1& 1& 1& 1& 1& 1&1\cr
S  & 1& 1& 1& 1& 1& 1& 1& 1& 1& 1& 1& 1&1\cr
e_1& 1& 1& 0& 1& 0& 1& 0& 0& 1& 0& 0& 0&0\cr
e_2& 1& 1& 1& 0& 0& 1& 0& 0& 1& 0& 0& 0&1\cr
e_3& 1& 1& 0& 0& 0& 1& 0& 0& 0& 0& 0& 1&1\cr
e_4& 1& 1& 1& 1& 1& 0& 1& 0& 0& 1& 1& 1&0\cr
e_5& 1& 1& 0& 0& 0& 1& 0& 0& 0& 1& 1& 0&1\cr
e_6& 1& 1& 0& 0& 0& 0& 0& 0& 1& 1& 0& 0&1\cr
b_1& 1& 0& 1& 1& 0& 0& 0& 1& 1& 0& 1& 1&0\cr
b_2& 1& 0& 0& 0& 0& 1& 1& 1& 0& 1& 0& 0&1\cr
z_1& 1& 1& 0& 0& 0& 1& 1& 0& 1& 0& 1& 1&1\cr
z_2& 1& 1& 0& 0& 1& 1& 0& 0& 1& 0& 1& 1&1\cr
\alpha& 1& 1& 0& 1& 1&0&1&1&1&0&0&1&1 }
\eeq
where we introduced the notation
$\cc{v_i}{v_j} = e^{i\pi (v_i|v_j)}$.

The gauge group of the model is obtained from the untwisted sector and the
$x$--sector \cite{xmap}, which arises from the vector combination
\beq
x=1+S+\sum_{i=1}^6e_i+z_1+z_2=\{ \bar{\psi}^{1,\cdots,5}, \bar{\eta}^{1,2,3}\}.
\label{xvector}
\eeq
The untwisted sector gives rise to space--time vector bosons
transforming under the group symmetry:
\beqn
{\rm observable} ~: &~~~~~~~~SO(6)\times SU(2)\times SU(2) \times U(1)^3 \nonumber\\
{\rm hidden}     ~: &~~SO(4)^2\times SO(8)~~~~             \nonumber
\eeqn
where the three $U(1)$ factors, $U(1)_{1,2,3}$, in the observable gauge group
are generated by the complex world--sheet fermions ${\bar\eta}^{1,2,3}$.
The $x$--sector produces space--time vector bosons transforming under
the untwisted group symmetry as:
\beq
(      4 ,2,1,  {1\over2},{1\over2},-{1\over2})+
({\bar 4},2,1, -{1\over2},-{1\over2},{1\over2}).
\label{xvectorbosons}
\eeq
Hence, the $SO(6)\times SU(2)\times U(1)$ group symmetry is enhanced to
$SU(6)$. The $U(1)$ combination which is embedded in $SU(6)$ is
given by
\beq
U(1)_6= U(1)_1+U(1)_2-U(1)_3
\label{u6}
\eeq
and the two orthogonal combinations are given by
\beqn
U(1)_1^\prime & = & 2U_1-U_2+U_3 \label{u1p}\\
U(1)_2^\prime & = & U_2+U_3.\label{u1a}
\eeqn
An important feature of this model, as compared to models in which
the observable gauge symmetry is not enhanced, is that the
$U(1)_6$ combination is automatically anomaly free. On the
other hand, of the two orthogonal combinations, one combination may,
in general, be anomalous.
Additional space--times
vector bosons may arise from the sectors \cite{fknr,fkr,acfkr}
\begin{equation}
\mathbf{G} =
\left\{ \begin{array}{cccccc}
z_1          ,&
z_2          ,&
z_1 + z_2    ,&
\alpha       ,&
\alpha + z_1 ,&
                \cr
\alpha + z_2 ,&
\alpha + z_1 + z_2,&
\alpha + x ,&
\alpha + x + z_1&
\end{array} \right\} \label{stvsectors}
\end{equation}
and may further enhance the four dimensional gauge group.
Here we impose the condition that the additional space--time vector
bosons arising from the sectors in (\ref{stvsectors})
are projected out.

The matter spectrum in the model arises from untwisted and twisted
sectors. As the basis vectors, eq. (\ref{basis}),
that generate the model are identical
to those that generate the Pati--Salam models of \cite{acfkr}
the sectors of the models are identical. The difference is that
in the $SU(6)\times SU(2)$ models the states from the different sectors
coalesce into representations of the enhanced group. Since the sector
producing the enhanced symmetry is the $x$--sector, it entails that
for every sector $\alpha$, which produces massless states that transform
under the untwisted group symmetry, there is a corresponding
sector $\alpha+x$, which completes the states to representations of
the enhanced $SU(6)\times SU(2)$ group. The enumeration of all
the sectors that produce massless states in these models, and the type
of states that they can a priori give rise to, was presented in ref.
\cite{acfkr}. The full massless spectrum of the string model
generated by the basis vectors and GGSO projection phases in eq.
(\ref{BigMatrix}) is shown in tables \ref{untwisted}, \ref{otwisted}
and \ref{htwisted}, where we define the vector combination
$b_3\equiv b_1+b_2+x$. In table \ref{untwisted} we list the untwisted
gauge and and matter multiplets. The untwisted matter multiplets
consist of the three pairs of $15+\overline{15}$. Therefore, there
is no net chirality arising from the untwisted sector, as expected
due to the $Z_2\times Z_2$ orbifold structure. Due to the symmetric
boundary conditions assigned to the internal world--sheet fermions that
correspond to the compactified six dimensional torus, the untwisted
$({\bar 6}, 2) \oplus (6,2)$ are projected out by the GGSO projections.

The twisted matter states generated in the string vacuum of eq.
(\ref{BigMatrix})
produce the needed spectrum for viable phenomenology. It contains three
chiral generation plus an additional generation and anti--generation
that can get a heavy mass along a flat direction. These pair of additional
states is an artifact of the fishing algorithm that was used to fish the
particular model presented here. However, models that do not contain
the additional states are phenomenologically viable provided that only
states in the $(15,1)$ and $(\overline{15},1)$ are used as heavy Higgs
representations. In that case the untwisted matter states give rise
to the heavy Higgs multiplets. Electroweak bi--doublets are
obtained in the model from the twisted sectors. The model additionally contains
$SU(6)\times SU(2)$ singlet states, some of which transform in non--trivial
representations of the hidden sector gauge group.

Exotic representations may arise in the string models in the
$(6,1)$, $({\bar 6},1)$ and $(1,2)$ representations of
$SU(6)\times SU(2)$. Such exotic states carry fractional
electric charge and are severely constrained by
experimental observations. They are generic in heterotic--string
models that preserve the canonical GUT embedding of the
weak hypercharge \cite{schellekens, ww}. All the exotic states
are projected in our model by the choice of
GGSO projection coefficients, eq. (\ref{BigMatrix})
that define the model, rather than by giving them mass
along flat direction in the effective low energy field theory
\cite{fc}.



\renewcommand{\arraystretch}{1.25}

\scriptsize
\begin{table}
\begin{center}
  \begin{tabularx}{\textwidth}{|X|c|c|c|c|c|}
    \cline{1-6}
    \textbf{Sector} & \textbf{Field} & \textbf{ $SU(6) \times SU(2)_R $ } & \textbf{$U(1)'_1$} & \textbf{$ U(1)_A$} & \textbf{$SO(4)_1 \times SO(4)_2 \times SO(8)$} \\ \cline{1-6}
    \multirow{18}{*}{$ S \oplus S+x $}
    & $F_5$ & $ (\,15,1\,) $ & $-2$ & ~~0 & (\,1,1,1\,) \\   
    & $\bar{F}_5$ & $ (\,\overline{15},1\,) $ & ~~2 & ~~0 & (\,1,1,1\,) \\   
    & $F_6$ & $ (\,15,1\,) $       & ~~1 & $-1$ & (\,1,1,1\,) \\   
    & $\bar{F}_6$ & $ (\,\overline{15},1\,) $ & $-1$ & ~~1 & (\,1,1,1\,) \\   
    & $F_7$ & $ (\,15,1\,) $       & ~~1 & ~~1 & (\,1,1,1\,) \\   
    & $\bar{F}_7$ & $ (\,\overline{15},1\,) $ & $-1$ & $-1$ & (\,1,1,1\,) \\   
    &${{\Phi_{12}}}$& $ (\,1,1\,) $       & ~~0 & ~~2 & (\,1,1,1\,) \\   
    &${{\bar{\Phi}_{12}}}$& $ (\,1,1\,) $ & ~~0 & $-2$ & (\,1,1,1\,) \\   
    &${{\Phi_{34}}}$& $ (\,1,1\,) $       & ~~3 & ~~1 & (\,1,1,1\,) \\   
    &${{\bar{\Phi}_{34}}}$& $ (\,1,1\,) $ & $-3$ & $-1$ & (\,1,1,1\,) \\   
    &${{\Phi_{56}}}$& $ (\,1,1\,) $       & ~~3 & $-1$ & (\,1,1,1\,) \\   
    &${{\bar{\Phi}_{56}}}$& $ (\,1,1\,) $ & $-3$& ~~1 & (\,1,1,1\,) \\   
    &$\hskip1.6cm {{\Phi_{I}}},I=1,\dots,6$& $ (\,1,1\,) $ & ~~0 & ~~0 & (\,1,1,1\,) \\   
    \cline{1-6}
 \end{tabularx}
\caption{\it Untwisted
matter spectrum. }
\label{untwisted}
\end{center}
\end{table}

\bigskip
\normalsize

\scriptsize
\begin{table}
\begin{center}
 \begin{tabularx}{\textwidth}{|c|c|c|c|c|c|X}
    \cline{1-6}
    \textbf{Sector} & \textbf{Field} & \textbf{ $SU(6) \times SU(2)_R $ } & \textbf{$U(1)'_1$} & \textbf{$ U(1)_A $} & \textbf{$SO(4)_1 \times SO(4)_2 \times SO(8)$} \\ \cline{1-6}
    \multirow{6}{*}{\begin{tabular}{c}$ S+b_1+e_4$ \\ $\oplus$ \\ $S+b_1+e_4+x $\\ \\ \\ \end{tabular}}
    & $F_1$     & $ (\,15,1\,) $ & $~~1$ & $~~0$ & $(\,1,1,1\,)$ \\   
    & $\chi_{1}$ & $ (\,1,1\,) $ & $-3$  & $~~0$ & $(\,1,1,1\,)$ \\   
    & ${{\zeta_a}},a=1,2$ & $ (\,1,1\,) $ & $~~0$ & $~~1$ & $(\,1,1,1\,)$ \\   
    & ${{\bar{\zeta}_a}},a=1,2$ & $ (\,1,1\,) $ & $~~0$ & $-1$ & $(\,1,1,1\,)$ \\   \cline{1-6}
    \begin{tabular}{c}$ S+b_1+e_4+e_6$ \\ $\oplus$ \\ $S+b_1+e_4+e_6+x $ \end{tabular}
    & ${{\bar{f}_1}}$ & $ (\,\overline{6},2\,) $ & $~~1$ & $~~0$ & $(\,1,1,1\,)$ \\   \cline{1-6}
    \multirow{6}{*}{\begin{tabular}{c}$ S+b_2+e_2$ \\ $\oplus$ \\ $S+b_2+e_2+x $\\ \\ \\ \end{tabular}}
    & ${{F_2}}$    & $ (\,15,1\,) $ & $-\frac{1}{2}$ & $~~\frac{1}{2}$ & $(\,1,1,1\,)$ \\   
    & ${{\chi_2}}$ & $ (\,1,1\,) $ & $~~\frac{3}{2}$ & $-\frac{3}{2}$ & $(\,1,1,1\,)$ \\   
    & ${{\zeta_a}},a=3,4$ & $ (\,1,1\,) $ & $~~\frac{3}{2}$ & $~~\frac{1}{2}$ & $(\,1,1,1\,)$ \\   
    & ${{\bar{\zeta}_a}},a=3,4$ & $ (\,1,1\,) $ & $-\frac{3}{2}$ & $-\frac{1}{2}$ & $(\,1,1,1\,)$ \\\cline{1-6}
    \begin{tabular}{c}$ S+b_2+e_2+e_6$ \\ $\oplus$ \\ $S+b_2+e_2+e_6+x $ \end{tabular}
    & ${{\bar{f}_2}}$ & $ (\,\overline{6},2\,) $ & $-\frac{1}{2}$ & $~~\frac{1}{2}$ & $(\,1,1,1\,)$ \\   \cline{1-6}
    \multirow{6}{*}{\begin{tabular}{c}$ S+b_2+e_1+e_6$ \\ $\oplus$ \\ $S+b_2+e_1+e_6+x $ \\ \\ \\ \end{tabular}}
    & ${{F_3}}$ & $ (\,15,1\,) $ & $-\frac{1}{2}$ & $~~\frac{1}{2}$ & $(\,1,1,1\,)$ \\   
    & ${{\chi_3}}$ & $ (\,1,1\,) $ & $~~\frac{3}{2}$ & $-\frac{3}{2}$ & $(\,1,1,1\,)$ \\   
    & ${{\zeta_a}},a=5,6$ & $ (\,1,1\,) $ & $~~\frac{3}{2}$ & $~~\frac{1}{2}$ & $(\,1,1,1\,)$ \\   
    & ${{\bar{\zeta}_a}},a=5,6$ & $ (\,1,1\,) $ & $-\frac{3}{2}$ & $-\frac{1}{2}$ & $(\,1,1,1\,)$ \\ \cline{1-6}
    \begin{tabular}{c}$ S+b_2+e_1$ \\ $\oplus$ \\ $S+b_2+e_1+x $ \end{tabular}
    & ${{\bar{f}_3}}$ & $ (\,\overline{6},2\,) $ & $-\frac{1}{2}$ & $~\frac{1}{2}$ & $(\,1,1,1\,)$ \\   \cline{1-6}
    \multirow{6}{*}{\begin{tabular}{c}$ S+b_3+e_4+e_2$ \\ $\oplus$ \\ $S+b_3+e_4+e_2+x $ \\ \\ \\ \end{tabular}}
    & ${{F_4}}$ & $ (\,15,1\,) $   & $-\frac{1}{2}$ &  $-\frac{1}{2}$ & $(\,1,1,1\,)$ \\   
    & ${{\chi_4}}$ & $ (\,1,1\,) $ & $+\frac{3}{2}$ & $+\frac{3}{2}$  & $(\,1,1,1\,)$ \\   
    & ${{\zeta_a}},a=7,8$ & $ (\,1,1\,) $  & $~~\frac{3}{2}$ & $-\frac{1}{2}$  & $(\,1,1,1\,)$ \\   
    & ${{\bar{\zeta}_a}},a=7,8$ & $ (\,1,1\,) $  & $-\frac{3}{2}$ & $~~\frac{1}{2}$  & $(\,1,1,1\,)$ \\ \cline{1-6}
    \multirow{6}{*}{\begin{tabular}{c}$ S+b_3+e_4+e_2+e_1$ \\ $\oplus$ \\ $S+b_3+e_4+e_2+e_1+x $ \\ \\ \\ \end{tabular}}
    & ${{\bar{F}_4}}$ & $ (\,\overline{15},1\,) $ & $~~\frac{1}{2}$ & $~~\frac{1}{2}$ & $(\,1,1,1\,)$ \\   
    & ${{\chi_5}}$ & $ (\,1,1\,) $ & $-\frac{3}{2}$ & $-\frac{3}{2}$ & $(\,1,1,1\,)$ \\   
    & ${{\zeta_a}},a=9,10$ & $ (\,1,1\,) $ & $~~\frac{3}{2}$  & $-\frac{1}{2}$ & $(\,1,1,1\,)$ \\   
    & ${{\bar{\zeta}_{a}}},a=9,10$ & $ (\,1,1\,) $ & $-\frac{3}{2}$ & $~~\frac{1}{2}$  & $(\,1,1,1\,)$ \\\cline{1-6}
    \begin{tabular}{c}$ S+b_3+e_4$ \\ $\oplus$ \\ $S+b_3+e_4+x $ \end{tabular}
    & ${{f_1}}$ & $ (\,6,2\,) $ & $~~\frac{1}{2}$ & $~~\frac{1}{2}$ & $(\,1,1,1\,)$ \\   \cline{1-6}
    \begin{tabular}{c}$ S+b_3+e_4+e_1$ \\ $\oplus$ \\ $S+b_3+e_4+e_1+x $ \end{tabular}
    & ${{\bar{f}_4}}$ & $ (\,\overline{6},2\,) $ & $-\frac{1}{2}$ & $-\frac{1}{2}$ & $(\,1,1,1\,)$ \\   \cline{1-6}
    \multirow{2}{*}{$ S+b_2+x+e_1+e_2  $}
    & ${{\zeta_{11}}}$ & $ (\,1,1\,) $ & $~~\frac{3}{2}$ & $~~\frac{1}{2}$ & (\,1,1,1\,) \\   
    & ${{\bar{\zeta}_{11}}}$ & $ (\,1,1\,) $ & $-\frac{3}{2}$ & $-\frac{1}{2}$ & (\,1,1,1\,) \\   \cline{1-6}
    \multirow{2}{*}{$ S+b_2+x+e_6 $}
    & ${{\zeta_{12}}}$ & $ (\,1,1\,) $ & $~~\frac{3}{2}$ & $~~\frac{1}{2}$ & (\,1,1,1\,) \\   
    & ${{\bar{\zeta}_{12}}}$ & $ (\,1,1\,) $ & $-\frac{3}{2}$ & $-\frac{1}{2}$ & (\,1,1,1\,) \\   \cline{1-6}
    \end{tabularx}
    \caption{Observable twisted matter spectrum.}
    \label{otwisted}
\end{center}
\end{table}

\bigskip
\normalsize

\scriptsize

\renewcommand{\arraystretch}{1.5}

\begin{table}
\begin{center}
  \begin{tabularx}{\textwidth}{|X|c|c|c|c|c|}
    \cline{1-6}
    \textbf{Sector} & \textbf{Field} & \textbf{ $SU(6) \times SU(2)_R $ } & \textbf{$U(1)'_1$} & \textbf{$ U(1)_A $} & \textbf{$SO(4)_1 \times SO(4)_2 \times SO(8)$} \\ \cline{1-6}
    $ S+b_1+x+e_4+e_5  $     & $H_{11}$ & $ (\,1,1\,) $ & ~~0 & ~~1 & (\,4,1,1\,) \\   
    $ S+b_1+x+e_4+e_5+e_6  $ & $H_{21}$ & $ (\,1,1\,) $ & ~~0 & ~~1 & (\,1,4,1\,) \\   
    $ S+b_1+x+e_3+e_4+e_6  $ & $Z_1$ & $ (\,1,1\,) $    & ~~0  & ~~1 & $(\,1,1,8_v\,)$ \\   \cline{1-6}
    $ S+b_2+x+e_1+e_2+e_5  $ & $H_{12}$ & $ (\,1,1\,) $ & $~~\frac{3}{2}$ & $~~\frac{1}{2}$ & (\,4,1,1\,) \\   
    $ S+b_2+x+e_1+e_2+e_5+e_6  $ & $H_{22}$ & $ (\,1,1\,) $ & $~~\frac{3}{2}$ & $~~\frac{1}{2}$ & (\,1,4,1\,) \\   
    $ S+b_2+x+e_5+e_6  $ & $H_{13}$ & $ (\,1,1\,) $ & $~~\frac{3}{2}$ & $~~\frac{1}{2}$ & (\,4,1,1\,) \\   
    $ S+b_2+x+e_5  $ & $H_{23}$ & $ (\,1,1\,) $ & $~~\frac{3}{2}$ & $~~\frac{1}{2}$ & (\,1,4,1\,) \\   \cline{1-6}
    $ S+b_3+x+e_1+e_3+e_4  $ & $Z_2$ & $ (\,1,1\,) $ & $-\frac{3}{2}$ & $~~\frac{1}{2}$ & $(\,1,1,8_v\,)$ \\   
    $ S+b_3+x+e_3+e_4  $ & $Z_3$ & $ (\,1,1\,) $ & $~~\frac{3}{2}$ & $-\frac{1}{2}$ & $(\,1,1,8_v\,)$ \\   \cline{1-6}
    $ S+b_1+x+z_2+e_3+e_4+e_5+e_6 $ & $Z_4$ & $ (\,1,1\,) $ & ~~0 & ~~1 & $(\,1,1,8_s'\,)$ \\   
    $ S+b_1+x+z_2+e_4+e_5  $ & $Z_5$ & $ (\,1,1\,) $ & ~~0 & ~~1 & $(\,1,1,8_s\,)$ \\   \cline{1-6}
    $ S+b_1+x+z_1+e_3+e_4+e_5+e_6  $ & $H_{121}$ & $ (\,1,1\,) $ & ~~0 & ~~1 & $(\,\overline{2},\overline{2},1\,)$ \\   
    $ S+b_1+x+z_1+e_3+e_4+e_5  $ & $H_{122}$ & $ (\,1,1\,) $ & ~~0 & ~~1 & $(\,2,2,1\,)$ \\   
    $ S+b_1+x+z_1+e_3+e_4+e_6  $ & $H_{123}$ & $ (\,1,1\,) $ & ~~0 & $-1$ & $(\,2,\overline{2},1\,)$ \\   
    $ S+b_1+x+z_1+e_3+e_4  $ & $H_{124}$ & $ (\,1,1\,) $ & ~~0 & $-1$ & $(\,\overline{2},2,1\,)$ \\   \cline{1-6}
    $ S+b_3+x+z_1  $ &$ H_{125}$ & $ (\,1,1\,) $ & $-\frac{3}{2}$ & $~~\frac{1}{2}$ & $(\,\overline{2},\overline{2},1\,)$ \\   
    $ S+b_3+x+z_1+e_2  $ & $H_{126}$ & $ (\,1,1\,) $ & $-\frac{3}{2}$ & $~~\frac{1}{2}$ & $(\,2,2,1\,)$ \\   
    $ S+b_3+x+z_1+e_1  $ & $H_{127}$ & $ (\,1,1\,) $ & $-\frac{3}{2}$ & $~~\frac{1}{2}$ & $(\,\overline{2},\overline{2},1\,)$ \\   
    $ S+b_3+x+z_1+e_2+e_1  $ & $H_{128}$ & $ (\,1,1\,) $ & $-\frac{3}{2}$ & $~~\frac{1}{2}$ & $(\,2,2,1\,)$ \\   \cline{1-6}
    \cline{1-6}
 \end{tabularx}
 \caption{Twisted hidden matter spectrum. }
 \label{htwisted}
\end{center}
\end{table}

\normalsize

\section{The superpotential }\label{pheno}

Using the methodology of ref. \cite{kln}
for the calculation of renormalisable and
nonrenormalisable terms, we calculate the cubic level
superpotential of our $SU(6)\times SU(2)$ string model.
The trilevel superpotential is given by

\begin{equation}
\begin{aligned}
& \frac{W_{SM}}{g\sqrt{2}}=
\ \mbox{}\hphantom{+}F_1\,{{F_2}}\,{{F_4}}+
   {{\bar{f}_1}}\,{{\bar{f}_2}}\,{{F_4}}\,+ \,{{\bar{f}_3}}\,{{\bar{f}_4}}\,F_{1}+
    {{\bar{f}_4}}\,f_{1}\,{{F_3}}\, + \,F_1\,F_{1}\,F_{5}+{{F_2}}\,{{F_2}}\,F_{6}\\
&\mbox{}+{{F_4}}\,{{F_4}}\,F_{7} + {{\bar{f}_1}}\,{\bar f}_{1}\,F_{5}+
   {{\bar{f}_2}}\,{{\bar{f}_2}}\,F_{6}+
   {{\bar{f}_4}}\,{{\bar{f}_4}}\,F_{7}\, + \,{{F_3}}\,{{F_3}}\,F_{5}+
   {{\bar{f}_3}}\,{{\bar{f}_3}}\,F_{5}\\
&+{{f_1}}\,{{f_1}}\,\bar{F}_{7}+{{\bar{F}_4}}\,{{\bar{F}_4}}\,\bar{F}_{7}\, +
   \bar{F}_{5}\,F_1\,\chi_{1}+\bar{F}_{6}\,{{F_2}}\,{{\chi_2}}+
   \bar{F}_{6}\,{{F_3}}\,{{\chi_3}}+\bar{F}_{7}\,{{F_4}}\,{{\chi_4}}\\
&+F_{7}\,{{\bar{F}_4}}\,{{\chi_5}}\,+
   \chi_{1}\,{{\chi_2}}\,{{\chi_4}} + \,F_5\,F_6\,F_7+
   \bar{F}_5\,\bar{F}_6\,\bar{F}_7\, +
   \,{{\Phi_{12}}}\,{{\bar{\Phi}_{34}}}\,{{\Phi_{56}}}+
   {{\bar{\Phi}_{12}}}\,{{\Phi_{34}}}\,{{\bar{\Phi}_{56}}}\,\\
&+\,\bar{F}_5\,F_6\,{{\bar{\Phi}_{56}}}+\bar{F}_5\,F_7\,{{\bar{\Phi}_{34}}} +
   F_5\,\bar{F}_6\,{{\Phi_{56}}}+F_5\,\bar{F}_7\,{{\Phi_{34}}}+
   \bar{F}_6\,F_7\,{{\bar{\Phi}_{12}}}+\bar{F}_7\,F_6\,{{\Phi_{12}}}\,\\
&+\frac{1}{\sqrt{2}}\left\{\,F_1\,{{\bar{F}_4}}\,{{\bar{\zeta}_{11}}}+
   {{\bar{f}_1}}\,{{f_1}}\,{{\bar{\zeta}_{12}}}\right\}
+\,\left\{{{\zeta_2}}\,{{\bar{\zeta}_7}}+{{\zeta_1}}\,{{\bar{\zeta}_8}}\right\}\,{{\chi_2}} +\left\{{{\bar{\zeta}_1}}\,{{\bar{\zeta}_3}}+{{\bar{\zeta}_2}}\,{{\bar{\zeta}_4}}\,\right\}{{\chi_4}}\\
&
+\left\{{{\zeta_3}}\,{{\zeta_7}}+{{\zeta_4}}\,{{\zeta_8}}+
{{{\zeta}_{10}}}\,{{\zeta_{11}}}\,\right\}\chi_{1}+{{\zeta_1}}\,{{\zeta_{11}}}\,{{\chi_5}}\,
+\frac{1}{\sqrt{2}}\left\{{{\zeta_1}}\,{{{\zeta}_{10}}}\,{{\bar{\zeta}_{11}}}+{{\zeta_2}}\,{{{\zeta}_9}}\,{{\bar{\zeta}_{11}}}\,\right\}  \\
&+\left\{\,{{\zeta_1}}\,{{\zeta_1}}+{{\zeta_2}}\,{{\zeta_2}}\,\right\}\,\bar{\Phi}_{12} +\,\left\{{{\bar{\zeta}_1}}\,{{\bar{\zeta}_1}}+{{\bar{\zeta}_2}}\,{{\bar{\zeta}_2}}\,\right\}\,\Phi_{12} \\
&+\left\{\strut\,{{\zeta_3}}\,{{\zeta_3}}+{{\zeta_4}}\,{{\zeta_4}}+{{\zeta_5}}\,{{\zeta_5}}+{{\zeta_6}}\,{{\zeta_6}}+{{\zeta_{11}}}\,{{\zeta_{11}}}+{{\zeta_{12}}}\,{{\zeta_{12}}}\,\right\}\,\bar{\Phi}_{34}+\left\{\,{{\bar{\zeta}_3}}\,{{\bar{\zeta}_3}}+{{\bar{\zeta}_4}}\,{{\bar{\zeta}_4}}\right. \\
&+\left.\strut{{\bar{\zeta}_5}}\,{{\bar{\zeta}_5}}+{{\bar{\zeta}_6}}\,{{\bar{\zeta}_6}}+{{\bar{\zeta}_{11}}}\,{{\bar{\zeta}_{11}}}+{{\bar{\zeta}_{12}}}\,{{\bar{\zeta}_{12}}}\,\right\}\,\Phi_{34}+\{\,{{\zeta_8}}\,{{\zeta_8}}+{{\zeta_7}}\,{{\zeta_7}}+{{{\zeta}_{10}}}\,{{{\zeta}_{10}}}+{{{\zeta}_9}}\,{{{\zeta}_9}}\,\}\,\bar{\Phi}_{56} \\
&+\{\,{{\bar{\zeta}_8}}\,{{\bar{\zeta}_8}}+{{\bar{\zeta}_7}}\,{{\bar{\zeta}_7}}+{{\bar{\zeta}_9}}\,{{\bar{\zeta}_9}}+{{\bar{\zeta}_{10}}}\,{{\bar{\zeta}_{10}}}\,\}\,\Phi_{56}+\{\,{{\zeta_{11}}}\,{{\bar{\zeta}_{11}}}+{{\zeta_{12}}}\,{{\bar{\zeta}_{12}}}\,\}\,{\Phi}_{4} \\
&+\left\{\strut\,H_{11}\,H_{11}+H_{21}\,H_{21}+H_{121}\,H_{121}+H_{122}\,H_{122}+Z_{1}\,Z_{1}+Z_{4}\,Z_{4}+Z_{5}\,Z_{5}\,\right\}\,{{\bar{\Phi}_{12}}}\\
&+\left\{\strut\,H_{12}\,H_{12} +H_{22}\,H_{22}+H_{13}\,H_{13}+H_{23}\,H_{23}\,\right\}\,{{\bar{\Phi}_{34}}} +Z_1\,Z_2\,{{\chi_3}} +\frac{1}{\sqrt{2}}\,Z_1\,Z_3\,{{\bar{\zeta}_{12}}} \\
&+\left\{\strut\,H_{125}\,H_{125}+H_{126}\,H_{126}+H_{127}\,H_{127}+H_{128}\,H_{128} +Z_{2}\,Z_{2}\,\right\}\,{{\Phi_{56}}} \\
&+\left\{\strut\,H_{123}\,H_{123}+H_{124}\,H_{124}\,\right\}\,{{\Phi_{12}}} +Z_{3}\,Z_{3}\,{{\bar{\Phi}_{56}}}+\left\{\,H_{11}\,H_{12}+H_{21}\,H_{22}\,\right\}\,{{\chi_5}}
\text{ .}
\end{aligned}
\label{superpot}
\end{equation}


The string vacuum contains one anomalous $U(1)$. In our model the anomalous
$U(1)$ combination is given in eq. (\ref{u1a}).
Hence, $U(1)_A\equiv U(1)_2^\prime$ and,
\begin{equation}
\nonumber \text{Tr}\,U(1)_A=72
\end{equation}
The anomalous $U(1)$ is a combination of the
$U(1)_{1,2,3}$. It is orthogonal to the combination of
$U(1)_{1,2,3}$ that is
embedded in $SU(6)$. Consistently, the family universal
$U(1)$ combination, which
is embedded in $SU(6)$, is anomaly free. This is an important
distinction in comparison to the FSU5, SLM and PS string models of
refs. \cite{fsu5}, \cite{slm} and \cite{alr}, respectively,
in which the family universal combination of $U(1)_{1,2,3}$ is anomalous.
Therefore, in these models the family universal combination must be broken
at the string scale and cannot remain unbroken down to low scales.
By contrast, the LRS models \cite {lrs}
produce models in which all three
$U(1)$s are anomaly free, whereas in the $SU(6)\times SU(2)$
models the family universal combination is anomaly free
by virtue of its embedding in $SU(6)$.
Of the two orthogonal combinations, given by eqs. (\ref{u1p}) and
(\ref{u1a}), the first is anomaly free, whereas the second is anomalous.

The anomalous $U(1)_A$ is broken by the Green--Schwarz--Dine--Seiberg--Witten
mechanism \cite{dsw} in which a potentially large Fayet--Iliopoulos
$D$--term $\xi$ is generated by the VEV of the dilaton field.
Such a $D$--term would, in general, break supersymmetry, unless
there is a direction $\hat\phi=\sum\alpha_i\phi_i$ in the scalar
potential for which $\sum Q_A^i\vert\alpha_i\vert^2<0$ and that
is $D$--flat with respect to all the non--anomalous gauge symmetries
along with $F$--flat. If such a direction
exists, it will acquire a VEV, cancelling the Fayet--Iliopoulos
$\xi$--term, restoring supersymmetry and stabilising the vacuum.
The $D$--term flatness constraints in our model, assuming zero VEVs for the hidden sector fields, are given by:

\begin{equation}
\begin{aligned}
&U(1)_1^{\prime} \text{: } D_1=\\
&-6|\chi_{1}|^2+3|{{\chi_2}}|^2+3|{{\chi_3}}|^2+3|{{\chi_4}}|^2-3|{{\chi_5}}|^2
  +3(|{{\zeta_4}}|^2+|{{\zeta_3}}|^2-|{{\bar{\zeta}_3}}|^2-|{{\bar{\zeta}_4}}|^2) \\
&+3(|{{\zeta_5}}|^2-|{{\bar{\zeta}_6}}|^2+|{{\zeta_6}}|^2-|{{\bar{\zeta}_5}}|^2)
+3(|{{\zeta_8}}|^2-|{{\bar{\zeta}_8}}|^2+|{{\zeta_7}}|^2-|{{\bar{\zeta}_7}}|^2) \\
&+3(|{{{\zeta}_9}}|^2+|{{{\zeta}_{10}}}|^2-|{{\bar{\zeta}_9}}|^2-|{{\bar{\zeta}_{10}}}|^2)
+3(|{{\zeta_{11}}}|^2-|{{\bar{\zeta}_{11}}}|^2+|{{\zeta_{12}}}|^2-|{{\bar{\zeta}_{12}}}|^2) \\
&+6(|\Phi_{34}|^2-|\bar{\Phi}_{34}|^2)+6(|\Phi_{56}|^2-6|\bar{\Phi}_{56}|^2) \\
&+2|\bar{f}_{1}|^2-|{{\bar{f}_2}}|^2-|{{\bar{f}_3}}|^2+|{{f_1}}|^2-|{{\bar{f}_4}}|^2\\
&  +2|F_{1}|^2-|{{F_2}}|^2-|{{F_3}}|^2-|{{F_4}}|^2+|{{\bar{F}_4}}|^2 \,=\,0 \text{ ,}
\end{aligned}
\end{equation}

\begin{equation}
\begin{aligned}
&U(1)_A \text{: } D_A=\\
&-3|{{\chi_2}}|^2-3|{{\chi_3}}|^2+3|{{\chi_4}}|^2-3|{{\chi_5}}|^2+
   2(|{{\zeta_1}}|^2+2|{{\zeta_2}}|^2-2|{{\bar{\zeta}_1}}|^2-2|{{\bar{\zeta}_2}}|^2) \\
&+(|{{\zeta_4}}|^2+|{{\zeta_3}}|^2-|{{\bar{\zeta}_3}}|^2-|{{\bar{\zeta}_4}}|^2)
  +(|{{\zeta_5}}|^2-|{{\bar{\zeta}_6}}|^2+|{{\zeta_6}}|^2-|{{\bar{\zeta}_5}}|^2) \\
&+(|{{\bar{\zeta}_7}}|^2+|{{\bar{\zeta}_8}}|^2-|{{\zeta_7}}|^2-|{{\zeta_8}}|^2)
  +(|{{\bar{\zeta}_9}}|^2+|{{\bar{\zeta}_{10}}}|^2-|{{{\zeta}_{10}}}|^2-|{{{\zeta}_9}}|^2)\\
&+(|{{\zeta_{11}}}|^2-|{{\bar{\zeta}_{11}}}|^2+|{{\zeta_{12}}}|^2-|{{\bar{\zeta}_{12}}}|^2) +
   4(|\Phi_{12}|^2-|\bar{\Phi}_{12}|^2)+2(|\Phi_{34}|^2-|\bar{\Phi}_{34}|^2) \\
&+2(|\bar{\Phi}_{56}|^2-|\Phi_{56}|^2)
  +  |{{\bar{f}_2}}|^2+|{{\bar{f}_3}}|^2+|{{f_1}}|^2-|{{\bar{f}_4}}|^2\\
&+ |{{F_2}}|^2+|{{F_3}}|^2-|{{F_4}}|^2+|{{\bar{F}_4}}|^2 \,
+\ \frac{3\,g_{\text{string}}^2\,M_P^2}{8\pi^2} \,
=\,0 \text{ .}
\end{aligned}
\label{du1a}
\end{equation}
In eq. (\ref{du1a}) $g$ is the gauge coupling in the effective field
theory. The set of $F$--flatness constraints is obtained by requiring
\begin{equation}
\left\langle
\frac{\partial W}{\partial \eta_i}\right\rangle=0 \text{ , }
~~\forall \eta_i \text{\ .}
\end{equation}

Let us briefly discuss the phenomenology of our model. The model's spectrum consists of $7\times\left(15,1\right)+4\times\left(\overline{15},1\right)$, namely
$F_1,F_2,F_3,F_4,{{F_5}},{{F_6}},F_7,\bar{F}_4,\bar{F}_5,\bar{F}_6,{{\bar{F}_7}}$, and  $4\times\left(\bar{6},2\right)+\left(6,2\right)$, namely ${{\bar{f}_1}},{{\bar{f}_2}},{{\bar{f}_3}},{{\bar{f}_4}},{{f_1}}$ and a number of $SU(6)\times{SU(2)}$ singlets. An important issue is that of the generation mass hierarchy. In $SU(6)\times{SU(2)}$ models all fermions in a generation
receive masses from a single invariant in the superpotential. Fermion mass hierarchy requires that only the heavy generation obtains mass term at leading order, whereas the
mass terms of the lighter generations are generated at higher orders. As mentioned earlier there are two possible embeddings the electroweak ${SU(2)}_L$
gauge symmetry in $SU(6)\times{SU(2)}$. Both  embeddings can be realised in our string model. Considering the $SU(6)\times{SU(2)}_R$ embedding of \eqref{fli151}-\eqref{fri62}
the breaking to the Standard Model gauge group can be realised with a set of $\left(15,1\right)+\left(\overline{15},1\right)$ and $\left(6,2\right)+\left(\bar{6},2\right)$
Higgs multiplets. The remaining three $\left(15,1\right)+\left(\overline{15},1\right)$ multiplets may become massive from cubic mass terms in the superpotential \eqref{superpot}.
There are three candidate fermion generation mass couplings in the superpotential ${{\bar{f}_1}}\,{{\bar{f}_2}}\,{{F_4}}+f_2\,{{\bar{f}_4}}\,F_1+{{\bar{f}_4}}\,{{\bar{f}_1}}\,{{F_3}}$. Assuming ${{\bar{f}_1}},{{f_1}}$ to play the
role of one of the heavy Higgs sets, this requires ${{\bar{\zeta}_{12}}}=0$. We are left with a single fermion generation mass term at tri-level $f_2\,{{\bar{f}_4}}\,F_1$. This is compatible
with the requirement that only one generation becomes massive at tri-level. Masses for the additional  $d$--quark like triplets, accommodated in $(15,1)$ and $(\overline{15},1)$
may arise from couplings of the form ${\cal F}_L^i{\cal F}_L^j{\cal H}$.
In the case of $SU(6)\times{SU(2)}_L$ embedding of \eqref{fri151}-\eqref{fli62}, the breaking to the Standard Model gauge group is realized by two sets of $\left(15,1\right)+\left(\overline{15},1\right)$ Higgs
multiplets. The additional two sets of  $\left(15,1\right)+\left(\overline{15},1\right)$ may receive masses from tri-level superpotential terms while the extra $\left(6,2\right)+\left(\bar{6},2\right)$ set can be removed by assigning a VEV to ${{\bar{\zeta}_{12}}}$ which provides mass to ${{\bar{f}_1}},{{f_1}}$. A single fermion family mass coupling
is obtained also in this scenario namely $f_2\,{{\bar{f}_4}}\,F_1$, consistent with the above mentioned phenomenological requirements.
Additional triplets may also acquire heavy masses through superpotential couplings of the type ${\cal F}_L^i{\cal F}_L^j{\cal H}$.

As the string model contains several pairs of
electroweak Higgs doublets, we need to ensure that at least one light pair survives in order to provide
masses to the chiral fermions.
The electroweak Higgs doublets
are accommodated in the $(6,2)+(\bar{6},2)$ representations of $SU(6)\times SU(2)$, and correspond to the fields ${{\bar{f}_1}}$, ${{\bar{f}_2}}$, ${{\bar{f}_3}}$, ${{\bar{f}_4}}$
and ${{f_1}}$ .

The Higgs doublets mass matrix is given by:
\begin{align}
	M_f=\begin{pmatrix}
F_5&{{F_4}}&0&{{\bar{\zeta}_{12}}}&{{F_3}}\\
{{F_4}}&F_6&0&0&0\\
0&0&F_6&0&F_1\\
{{\bar{\zeta}_{12}}}&0&0&\bar{F}_7&0\\
{{F_3}}&0&F_1&0&F_7
\end{pmatrix} \ .
\label{higgsmassmatrix}
\end{align}
The requirement of existence of light Higgs doublets dictates the vanishing of the  determinant of the Higgs mass matrix, {\it i.e.}
\begin{align}
	\det(M_f)=[(F_5F_6-{{F_4}}^{\ 2})\bar{F}_7-
{{\bar{\zeta}_{12}}}^{\ 2}F_6](F_6F_7-F_1^{\,2})-{{F_3}}^{\ 2}F_6^{\,2}\bar{F}_7 &=0\ .
\label{higgsdeterminant}
\end{align}

We now discuss an example for the flatness conditions,
taking the field assignment in eqs. (\ref{fri62}) and (\ref{fli151}),
for $SU(6)\times{SU(2)}_R$ scenario. In the solution that we discuss here we will assume that the $SU(6)\times
SU(2)$ symmetry is broken to the Standard Model at high scale.
The scenario with a light extra $U(1)$ requires first a more detailed
effective field theory analysis of the $SU(6)\times SU(2)$ GUT model.

The breaking of $SU(6)\times SU(2)$ is
obtained by choosing the pair of $(15,1)+(\overline{15},1)$, for example,
${{F_3}}$ and ${{\bar{F}_4}}$. The condition that $F_6\bar{F}_7=0$ ensures that
these fields are massless. To break the Pati-Salam group, we choose a
pair of $(6,2)+(\bar{6},2)$, for example ${{\bar{f}_2}}$ and ${{f_1}}$.
The minimal choice
\begin{align}
{{F_3}}&={{\bar{F}_1}}\\
{{\bar{f}_2}}&={{f_1}}\\
|{{\bar{\Phi}_{12}}}|^2&=
  +  \frac{1}{2}|{{\bar{f}_2}}|^2+\frac{1}{2}|{{F_3}}|^2
+\ \frac{3\,g_{\text{string}}^2\,M_P^2}{32\pi^2}
 \end{align}
with all the other field VEVs vanishing, satisfies all F and D-flatness conditions.

\section{Conclusions}\label{conclusion}

In this paper we studied the construction of heterotic--string models
with $SU(6)\times SU(2)$ GUT group.
As a GUT model this case as been scarcely studied in the literature.
As a string model it possesses
some interesting properties. In particular it gives rise to an additional
anomaly free flavour universal
$U(1)$ symmetry. While string models generically produce additional
$U(1)$ symmetries, it is
often found that the flavour universal combinations, beyond the $SU(10)$
group that gives rise
to the Standard Model subgroup, are anomalous. Such anomalous $U(1)$
symmetries must be broken
at the string scale, and cannot remain unbroken down to lower scales.
It is therefore of further
interest to explore in detail the phenomenology of the
$SU(6)\times SU(2)$ GUT models,
and to study whether the additional $U(1)$ can indeed
remain unbroken down to low scales
while satisfying all other phenomenological constraints.
The motivation to keeping
the $Z^\prime$ unbroken down to low scales stems from his
possible role in suppressing
proton decay mediating operators, as well as
that of the $\mu$--term.
In this regard, we remark that
while substantial amount of phenomenological
studies in the literature are devoted to string inspired $Z^\prime$ models,
the viability of a low scale $Z^\prime$ in a string derived model is yet to
be demonstrated.

In this paper we constructed a three generation heterotic--string with
$SU(6)\times SU(2)$
GUT group. We used our recently developed fishing algorithm to obtain
a model with pre--conditioned phenomenological properties.
By using the free fermionic formalism for the construction of
string compactifications, we generate a large space of
vacua and use statistical sampling to extract a model
with the desired characteristics\footnote{We note that analysis
of large sets of string vacua has also been carried
out by other groups \cite{statistical}.}.
The model contains the heavy and light Higgs representations needed to
break the gauge symmetry down to the Standard Model and to generate fermion masses.
We extracted the cubic level superpotential, which is compatible with all the
symmetries and the string selection rules. By analysing the superpotential
we showed that there exist supersymmetric $F$-- and $D$--flat directions
that produce a fermion mass term at leading order for a single
family. The mass terms of the lighter families, which must then arise
from higher order terms in the superpotential, are naturally suppressed
relative to the heavy family mass term. Additionally, due to pre--imposed
constraints in our statistical sampling procedure,
our model is free of
all massless exotic representations that carry fractional electric charge.
The work reported in this paper therefore enlarges the space of
phenomenologically viable string vacua to a new class of models. 

\section{Acknowledgements}

LB, IG and JR would like to thank the University of Liverpool, and 
AEF would like to thank CERN and Oxford University, for hospitality.
AEF is supported in part by STFC under contract ST/J000493/1.
JR work has been supported in part by
the ITN network ``UNILHC'' (PITN-GA-2009-237920).
This research has been co-financed by the European Union (European 
Social Fund \&\#8211; ESF) and Greek national funds through the 
Operational Program "Education and Lifelong Learning" of the National 
Strategic Reference Framework (NSRF) - Research Funding Program: 
Thales. Investing in knowledge society through the European Social Fund.



\bigskip
\medskip

\bibliographystyle{unsrt}

\begin{thebibliography}{99}

\bibitem{plg} A. Font, L.E. Ibanez and F. Quevedo, \PLB{228}{1989}{79};\\
              A.E. Faraggi and D.V. Nanopoulos, \MODA{6}{1991}{61};\\
              J. Pati, \PLB{388}{1996}{532};\\
              A.E. Faraggi, \PLB{499}{2001}{147};\\
              C. Coriano, M. Guzzi and A.E. Faraggi, \EJP{53}{2008}{421}.

\bibitem{fff}
I. Antoniadis, C. Bachas, and C. Kounnas, \NPB{289}{1987}{87};\\
H. Kawai, D.C. Lewellen, and S.H.-H. Tye, \NPB{288}{1987}{1};\\
I. Antoniadis and C. Bachas, \NPB{298}{1988}{586}.

\bibitem{fsu5} I. Antoniadis, J. Ellis, J. Hagelin and D.V. Nanopoulos,
            \PLB{231}{1989}{65}.

\bibitem{slm} A.E. Faraggi, D.V. Nanopoulos and K. Yuan,
                    \NPB{335}{1990}{347};\\
              A.E. Faraggi, \PLB{278}{1992}{131}; \NPB{387}{1992}{239};\\
              G.B. Cleaver, A.E. Faraggi and D.V. Nanopoulos,
            \PLB{455}{1999}{135}.

\bibitem{alr} I. Antoniadis. G.K. Leontaris and J. Rizos,
                                \PLB{245}{1990}{161};\\
              G.K. Leontaris and J. Rizos, \NPB{554}{1999}{3};\\
              K. Christodoulides, A.E. Faraggi and J. Rizos, \PLB{702}{2011}{81}.


\bibitem{lrs}
    G.B. Cleaver, A.E. Faraggi and C. Savage, \PRD{63}{2001}{066001};
    G.B. Cleaver, D.J Clements and A.E. Faraggi, \PRD{65}{2002}{106003};

\bibitem{su421} G.B. Cleaver, A.E. Faraggi and
                                      S.M.E. Nooij, \NPB{672}{2003}{64}.


\bibitem{z2z21}
A.E. Faraggi, \PLB{326}{1994}{62}; \PLB{544}{2002}{207};\\
P. Berglund {\it et.al}, \PLB{433}{1998}{269}; \IJMP{15}{2000}{1345};\\
R. Donagi and A.E. Faraggi, \NPB{694}{2004}{187}.

\bibitem{z2z22}
E. Kiritsis, C. Kounnas, P.M. Petropoulos and J. Rizos, hep-th/9605011;\\
E. Kiritsis, C. Kounnas, P.M. Petropoulos and J. Rizos, \NPB{483}{1997}{141};\\
E. Kiritsis and C. Kounnas, \NPB{503}{1997}{117};\\
A. Gregori and C. Kounnas, \NPB{560}{1999}{135}.

\bibitem{acfkr}
B. Assel, C. Christodoulides, A.E. Faraggi, C. Kounnas and J. Rizos
                             \PLB{683}{2010}{306}; \NPB{844}{2011}{365}.

\bibitem{viraf1} A.E. Faraggi and V. Mehta, \PRD{84}{2011}{086006};
                                         \PLB{703}{2011}{567}.

\bibitem{viraf2} A.E. Faraggi and V. Mehta, paper in preparation.

\bibitem{dh} S. Dimopoulos and L.J. Hall, \NPB{255}{1985}{633}.

\bibitem{rt} J. Rizos and K. Tamvakis, \PLB{414}{1997}{277}.

\bibitem{kim} J.E. Kim, \PRD{85}{2012}{015012}.

\bibitem{raby} A. Anandakrishnan and S. Raby, arXiv:1205:1228.

\bibitem{gkr} A. Gregori, C.~Kounnas and J.~Rizos, \NPB{549}{1999}{16}.

\bibitem{fknr} A.E. Faraggi, C. Kounnas, S.E.M. Nooij and J. Rizos,
        hep-th/0311058; \NPB{695}{2004}{41}.

\bibitem{fkr} A.E. Faraggi, C. Kounnas and J. Rizos,
                \PLB{648}{2007}{84};
                \NPB{774}{2007}{208};\\
       T. Catelin-Julian, A.E. Faraggi, C. Kounnas and J. Rizos,
            \NPB{812}{2009}{103}.

\bibitem{xmap} A.E. Faraggi, \NPB{407}{1993}{57};
                            \EJP{49}{2007}{803}.

\bibitem{schellekens}A.N. Schellekens, \PLB{237}{1990}{363}.

\bibitem{ww}X.G. Wen and E. Witten, \NPB{261}{1985}{651};\\
G. Athanasiu, J. Atick, M. Dine, and W. Fischler,
                    \PLB{214}{1988}{55}.

\bibitem{fc} A.E. Faraggi, \PRD{46}{1992}{3204}.

\bibitem{kln} S.Kalara, J.L. Lopez and D.V. Nanopoulos, \NPB{353}{1991}{650};\\
              J. Rizos and K. Tamvakis, \PLB{262}{1991}{227};\\
              \AEF, \NPB{487}{1997}{55}.


\bibitem{dsw}   M.B. Green and J.H. Schwarz, \PLB{149}{1984}{117};\\
        M. Dine, N. Seiberg and E. Witten, \NPB{289}{1987}{589};\\
        J. Atick, L. Dixon and A. Sen, \NPB{292}{1987}{109}.

\bibitem{statistical} See {\it e.g.}: D. Senechal, \PRD{39}{1989}{3717};\\
        K.R. Dienes, \PRL{65}{1990}{1979}; \PRD{73}{2006}{106010};\\
            M.R. Douglas, \JHEP{0305}{2003}{046};\\
        R. Blumenhagen \etal, \NPB{713}{2005}{83};\\
        F. Denef and M.R. Douglas, \JHEP{0405}{2004}{072};\\
        T.P.T. Dijkstra, L. Huiszoon and A.N. Schellekens,
            \NPB{710}{2005}{3};\\
          B.S. Acharya, F. Denef and R. Valadro, \JHEP{0506}{2005}{056};\\
    P. Anastasopoulos, T.P.T. Dijkstra, E. Kiritsis and A.N. Schellekens,
        \NPB{759}{2006}{83};\\
        M.R. Douglas and W. Taylor, \JHEP{0701}{2007}{031};\\
    K.R. Dienes, M. Lennek, D. Senechal and V. Wasnik,
            \PRD{75}{2007}{126005};\\
    O. Lebedev \etal, \PLB{645}{2007}{88};\\
    E. Kiritsis, M. Lennek and A.N. Schellekens
        \JHEP{0902}{2009}{030}.


\end{thebibliography}

\end{document}